\documentstyle[11pt,aaspp4]{article}

\def\al{$\alpha$}
\def\bet{$\beta$}
\def\amin{$^\prime$}

\def\apj{{ApJ}}
\def\apjs{{ApJS}}
\def\asec{$^{\prime\prime}$}

\def\deg{$^{\circ}$}

\def\e#1{ $\times$ 10$^{#1}$}
\def\etal{{et al. }}

\def\flux{ergs s$^{-1}$ cm$^{-2}$}

\def\kms    {~km~s$^{-1}$}

\def\lum{ergs s$^{-1}$}

\def\mnras{{MNRAS}}
\def\nat{{Nature}}
\def\pasj{{PASJ}}
\def\pasp{{PASP}}
\def\percm2{cm$^{-2}$}
\def\persec{s$^{-1}$}

\def\lax    {${_<\atop^{\sim}}$ }

\def\refindent{\par\noindent\parskip=2pt\hangindent=3pc\hangafter=1 }

\def\farcm{\hbox{$.\mkern-4mu^\prime$}}

\slugcomment{To appear in {\it The Astrophysical Journal}.}
\lefthead{Ho et al.}
\righthead{X-Ray Properties of NGC 4639}

\begin{document}

\title{X-ray Properties of the Weak Seyfert 1 Nucleus in NGC 4639}

\author{Luis C. Ho}

\affil{Carnegie Observatories, 813 Santa Barbara St., Pasadena, CA 91101-1292}

\author{Andrew Ptak}

\affil{Carnegie Mellon University, Dept. of Physics, Pittsburg, PA 15213}

\author{Yuichi Terashima}

\affil{NASA/GSFC, Code 662, Greenbelt, MD 20771}

\author{Hideyo Kunieda}
 
\affil{Department of Astrophysics, Nagoya University, Chikusa-ku, Nagoya 
464-01, Japan}

\author{Peter J. Serlemitsos and Tahir Yaqoob}
  
\affil{NASA/GSFC, Code 662, Greenbelt, MD  20771} 

\and 

\author{Anuradha P. Koratkar}
  
\affil{Space Telescope Science Institute, 3700 San Martin Dr., Baltimore, MD 
21218}

\begin{abstract}
Low-luminosity active galactic nuclei (AGNs), of which NGC 4639 is a good 
example, constitute an important but poorly understood constituent of the 
nearby galaxy population.  We obtained observations of NGC 4639 with 
{\it ASCA} in order to investigate its mildly active Seyfert 1 nucleus at hard 
X-ray energies.  Koratkar \etal (1995) have previously shown that the nucleus 
is a pointlike source in the {\it ROSAT} soft X-ray band.  We detected in the 
2--10 keV band a compact central source with a luminosity of 8.3\e{40} \lum\ 
(for an adopted distance of 25.1 Mpc).  Comparison of the {\it ASCA} data with 
archival data taken with the {\it Einstein} and {\it ROSAT} satellites shows 
that the nucleus varies on timescales of months to years.  The variability 
could be intrinsic, or it could be caused by variable absorption.  More rapid 
variability, on a timescale of $\sim$10$^4$~s, may be present in the 
{\it ASCA} data.  The spectrum from 0.5 to 10 keV is well described by a model 
consisting of a lightly absorbed ($N_{\rm H} = 7.3 \times 10^{20}$ cm$^{-2}$) 
power law with a photon index of $\Gamma$ = 1.68$\pm$0.12.  We find no 
evidence for significant emission from a thermal plasma; if present, it can 
account for no more than $\sim$25\% of the flux in the 0.5--2.0 keV band.  The 
limited photon statistics of our data do not allow us to place significant 
limits on the presence of iron~K emission.

Despite its low luminosity, the X-ray properties of the nucleus of NGC 4639 
appear quite normal compared to those of more luminous AGNs.  The strength of 
its broad H\al\ line follows the correlation between broad H\al\ luminosity 
and hard X-ray luminosity previous known for luminous objects.  Images taken 
with the {\it Hubble Space Telescope} detected the nucleus in the ultraviolet 
at a strength relative to the X-ray band which appears to be quite typical 
of that found in other AGNs.  NGC 5033, another low-luminosity Seyfert 1 
galaxy whose optical characteristics closely resemble those of NGC 4639, has 
also been recently studied with {\it ASCA}, and we highlight some of the 
similarities between these two objects.
\end{abstract}


\section{Background}
NGC 4639 is one of the weakest known Seyfert galaxies.  Its 
active nucleus was discovered by Filippenko \& Sargent (1986) during the 
course of an extensive optical spectroscopic survey of nearby galaxies 
(Filippenko \& Sargent 1985; Ho, Filippenko, \& Sargent 1995).   NGC 4639 is 
a moderately-inclined, bulge-dominated barred spiral; Sandage \& Tammann (1981) 
assign it a Hubble type of SBb, and de Vaucouleurs \etal (1991) favor a 
classification of SABbc.  The Cepheid-based distance of the galaxy has been 
determined by Sandage \etal (1996) to be 25.1 Mpc.  The optical spectrum of 
the nucleus exhibits prominent hydrogen Balmer emission lines with velocities
of $\sim$4000 \kms\ at full-width half-maximum (FWHM).   According to the 
criteria of Osterbrock (1981), the relative strengths of the broad Balmer 
lines formally qualify the nucleus as a Seyfert of type 1.0 (Ho \etal 1997b).  
By traditional standards, however, the nucleus of NGC 4639 is intrinsically 
quite weak.  Ho \etal (1997b) measure a luminosity of $\sim$1\e{40} \lum\ for 
the broad H\al\ emission line.  The pointlike nucleus, while visible in {\it 
Hubble Space Telescope (HST)} images taken at optical wavelengths (Calvani 
\etal, in preparation), is nonetheless rather faint: $B\,\approx$ 19.5 mag, or 
$M_B\,\approx$ --12.5 mag ---  at least a factor of $10^4$ fainter than the 
most luminous Seyfert nuclei ($M_B\,\approx$ --23 mag; Weedman 1976).

Very little information exists for active galactic nuclei (AGNs) in 
the luminosity regime of NGC 4639.  Detailed studies of such objects 
are not only of intrinsic interest, but, by exploring new parameter 
space, may illuminate our understanding of the AGN phenomenon in general.  
The data deficiency is particularly acute at X-ray energies, where poor angular 
resolution and low sensitivity have been major hinderances in detecting and 
isolating a faint nucleus from the surrounding light of the host galaxy.  
Substantial progress has been made in the past few years with the advent of 
{\it ROSAT} and {\it ASCA} (see, e.g., Koratkar \etal 1995; Ishizaki \etal 
1996; Serlemitsos, Ptak, \& Yaqoob 1996; Iyomoto \etal 1998; Nicholson \etal 
1998; Ptak \etal 1999; Terashima 1999), but the samples studied so far remain 
very small and are largely biased against low-luminosity Seyferts.

Here we present new X-ray observations of NGC 4639 obtained with {\it ASCA}.  
We will combine these observations with existing data taken with {\it ROSAT} 
and with {\it HST} in order to derive a more complete picture of the X-ray 
properties of the source.  

\section{X-Ray Observations}
The {\it ASCA} data were acquired on 1997 December 17 and 23 UT; a description 
of the satellite can be found in Tanaka, Inoue, \&  Holt (1994).   {\it ASCA} 
consists of four identical X-ray telescopes whose focal plane is  equipped 
with two Solid-state Imaging Spectrometers (SIS0 and SIS1) and two Gas Imaging 
Spectrometers (GIS2 and GIS3).  The SIS covers the energy range 0.6--10.0 keV 
with a resolution (FWHM) of $E/\Delta E\,\approx 50$ at 6 keV and $\sim$20 at 
1.5 keV; each camera has a field-of-view of approximately 
22\amin$\times$22\amin, although in these observations only one CCD was active 
resulting in a field-of-view of $\sim$ 11\amin$\times$11\amin.  
The GIS has lower spectral resolution than the SIS ($E/\Delta E\,\approx 13$ 
at 6 keV and $\sim$7 at 1.5 keV), but it has higher efficiency in the hard 
X-rays over the bandpass 0.7--10.0 keV; the usable field-of-view of the GIS has 
a circular diameter $\sim$40\amin.  The point-spread function (PSF) of the 
images has a half-power diameter of 3\amin, with $\sim$20\% of the photons 
concentrated in a sharp core of diameter 1\amin.   In practice, the SIS 
delivers images of higher angular resolution than the GIS, whose PSF core is 
somewhat adversely affected by errors from position determination.

We operated the SIS in the 1-CCD faint mode, and the GIS was used 
in the nominal pulse-height mode.  We screened the data using standard 
criteria, which include exclusion of data taken when the elevation angle was 
less than 25\deg\ for SIS and less than 5\deg\ for GIS, when the geomagnetic 
cut-off rigidity was less than 6 GeV 
$c^{-1}$, and during passage through the South Atlantic Anomaly.  So-called 
hot and flickering pixels were also removed from the SIS data.  The final 
integration times for the SIS and GIS data sets are 66.8 ks and 71.1 ks, 
respectively.   The light curves and spectra, discussed in the next section, 
were extracted using a circular aperture of radius 4\amin\ for the SIS and 
6\amin\ for the GIS.  We measured the background from a source-free region 
within the same field.  After background subtraction, the count rates for 
SIS0, SIS1, GIS2, and GIS3, averaged over the two observations, are 
0.034, 0.033, 0.019, and 0.021 counts s$^{-1}$, respectively.  We combined 
the SIS0 and SIS1 spectra after appropriate gain corrections, and 
similarly for the GIS2 and GIS3 spectra.  To permit $\chi^2$ analysis, the 
spectra were binned such that each bin contains at least 20 counts.

\section{Analysis}

\subsection{Spatial Fitting}

\begin{figure}
\plotone{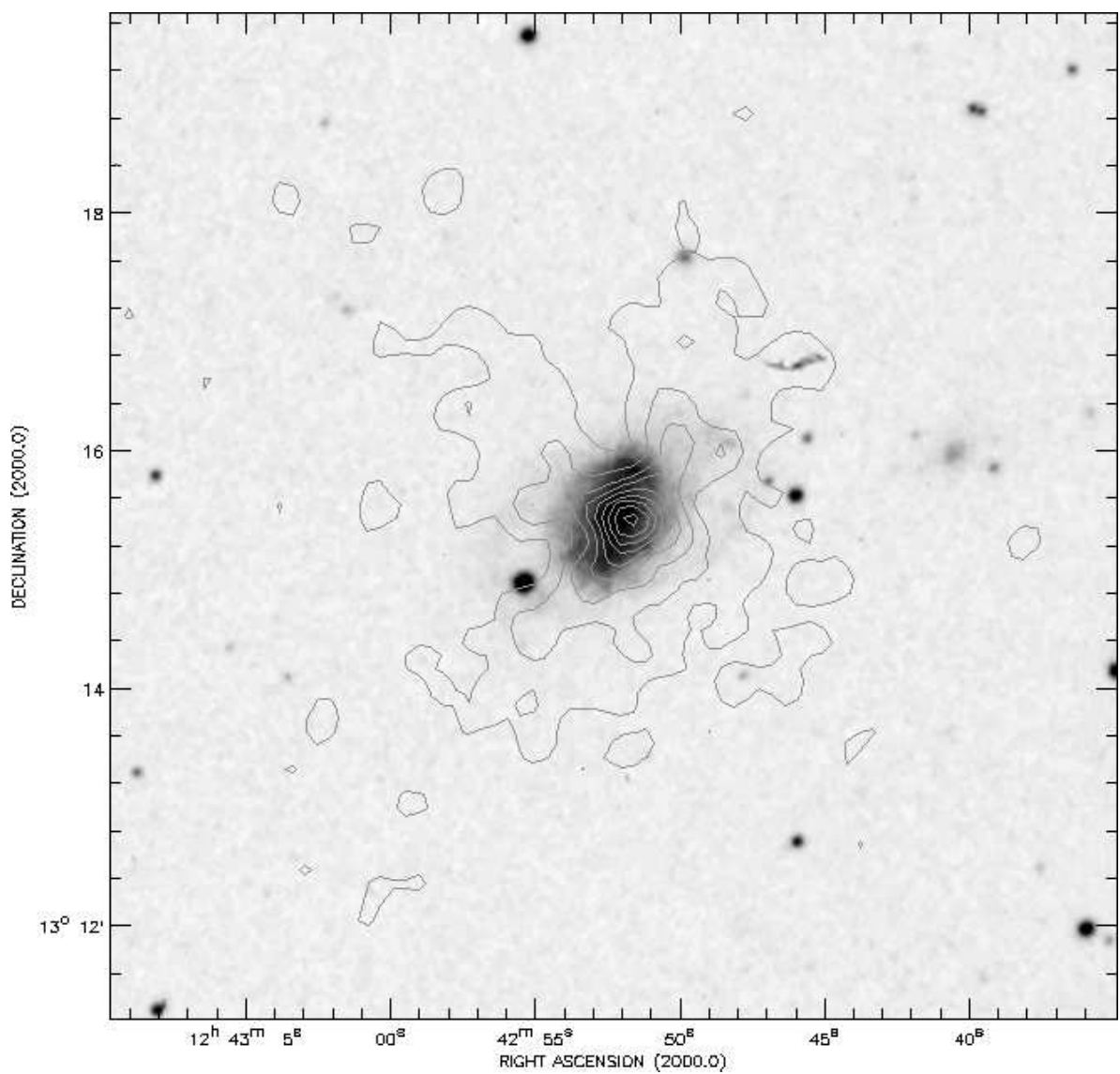}
\caption{{\it ASCA} SIS0+SIS1 contour image of NGC 4639 superposed on an 
optical image of the galaxy taken from the Digital Sky Survey. For the 
purposes of the display, the {\it ASCA} image has been smoothed with a 
Gaussian of width 6\asec.  The contour levels are linear, and they begin 
at 1.6\e{-5} counts s$^{-1}$.}
\end{figure}

Figure 1 displays the SIS0+SIS1 contour image of NGC 4639 taken on 1997 
December 17, superposed on an optical image of the galaxy taken from the 
Digital Sky Survey.  A compact source positionally coincides with the optical 
nucleus.  In order to properly analyze the spatial extent of X-ray data, 
particularly in the case of {\it ASCA}, the contribution of the PSF
must be taken into account.  In order to do this, we produced a 
model PSF and then convolved it with a model consisting of a single Gaussian, 
$S(r) = N{\rm exp}(-\frac{r^2}{2\sigma^2})$, or a double Gaussian,
$S(r) = N_1{\rm exp}(-\frac{r^2}{2\sigma_1^2}) + 
N_2{\rm exp}(-\frac{r^2}{2\sigma_2^2})$.  The best-fitting spectral model for 
the {\it ASCA} data (\S\ 3.2) was used to simulate the energy dependence of 
the PSF.  We only analyzed the SIS0 data from the first observation; combining 
data from both SIS detectors for spatial analysis is problematic because the 
source lies at different off-axis angles in the two detectors.  We find that a 
single Gaussian provides an acceptable fit: the source is consistent with 
being unresolved, with an upper limit size of $\sigma\,\approx$ 30\asec.  

We have also reanalyzed the {\it ROSAT} High-Resolution Imager (HRI) image of 
NGC 4639 previously published by Koratkar \etal (1995).  We adopt as a model 
of the HRI PSF the analytic formula given in 
{\tt http://heasarc.gsfc.nasa.gov/docs/rosat/faqs/hri\_psf\_faq1.html}.
The HRI data were split into two observations separated by $\sim$5 months.  
The latter observation is consistent with a single-Gaussian fit with 
$\sigma\,\approx$ 2\asec.  In addition to the dominant compact source, the 
earlier of the two observations shows measurable excess emission at 
$\sim$25\asec, at a level of $\sim$2\% of the total count rate.  This extra 
component may be due to a variable source, such as a background quasar, 
although inspection of the two images does not reveal any obvious point 
source apart from the galaxy nucleus.  Alternatively, the low-level emission 
may be an artifact of PSF variations, which have been known to occur at 
this level (E. Colbert, private communications).   Given that NGC 4639 does 
show evidence of long-term variability (\S\ 3.3), it is reasonable to assume 
that essentially all of the X-ray flux from NGC 4639 is unresolved, with a 
half-light radius on the order of 2\asec\ ($\sim 240$ pc) or less.

\subsection{Spectral Fitting}

After reducing the {\it ASCA} data, we extracted separate spectra for each of 
the SIS and GIS detectors, resulting in a set of eight source and eight 
background spectra for the two observations.  We find that the SIS0 and SIS1
spectra are consistent with each other, as are the GIS2 and GIS3 spectra, and, 
accordingly, the data from each detector were combined. (SIS and GIS spectra 
cannot be combined because the detectors have different bandpasses and 
responses.)  This resulted in two SIS and two GIS spectra.  In the following, 
we fit the four spectra simultaneously, leaving the overall normalization to 
be a free parameter (to allow for systematic uncertainty in the flux 
calibration of each set of detectors and for variability between the two 
observations).  A model consisting of a power law modified by photoelectric 
absorption provides an acceptable fit to all four spectra (the reduced 
$\chi^2$, $\chi^2_{\nu}$, is 0.98 for 491 degrees of freedom), resulting in 
a column density of $N_{\rm H} = 7.3 (2.2-12.9) \times 10^{20}$ cm$^{-2}$ and 
a photon index of $\Gamma$ = 1.68 (1.56--1.80), where the values in 
parentheses give the 90\% confidence interval assuming two interesting 
parameters.  For comparison, the foreground absorbing column due to the Galaxy 
is $N_{\rm H} = 2.3 \times 10^{20}$ cm$^{-2}$ (Murphy \etal 1996), and, for an 
assumed Case B$^{\prime}$ ratio H\al/H\bet\ = 3.1 (see Gaskell \& Ferland 1984),
the decrement of the narrow Balmer lines measured by Ho \etal (1997a) 
indicates an internal reddening of $E(B-V)\,\approx$ 0.03 mag, which 
corresponds to $N_{\rm H}\, \approx 2 \times 10^{20}$ cm$^{-2}$ for the 
conversion $E(B-V)\,=\,N_{\rm H}$/(5.8\e{21} cm$^{-2}$) mag (Bohlin, Savage, 
\& Drake 1978).  The spectra, their fits, and the residuals of the fits are 
shown in Figure 2.  The observed 2--10 keV flux is $1.1 \times 10^{-12}$ \flux\ 
(corresponding to a luminosity of 8.3\e{40} \lum\ for an adopted distance of 
25.1 Mpc) for all but the second SIS observation, which gave a flux $\sim$25\% 
higher.  {\it ASCA} fluxes typically have uncertainties of $\sim$10\%--20\%, 
so the flux discrepancy of the second observation is somewhat large but 
probably not inconsistent with the calibration uncertainty of the detectors.  

\begin{figure}
\plotone{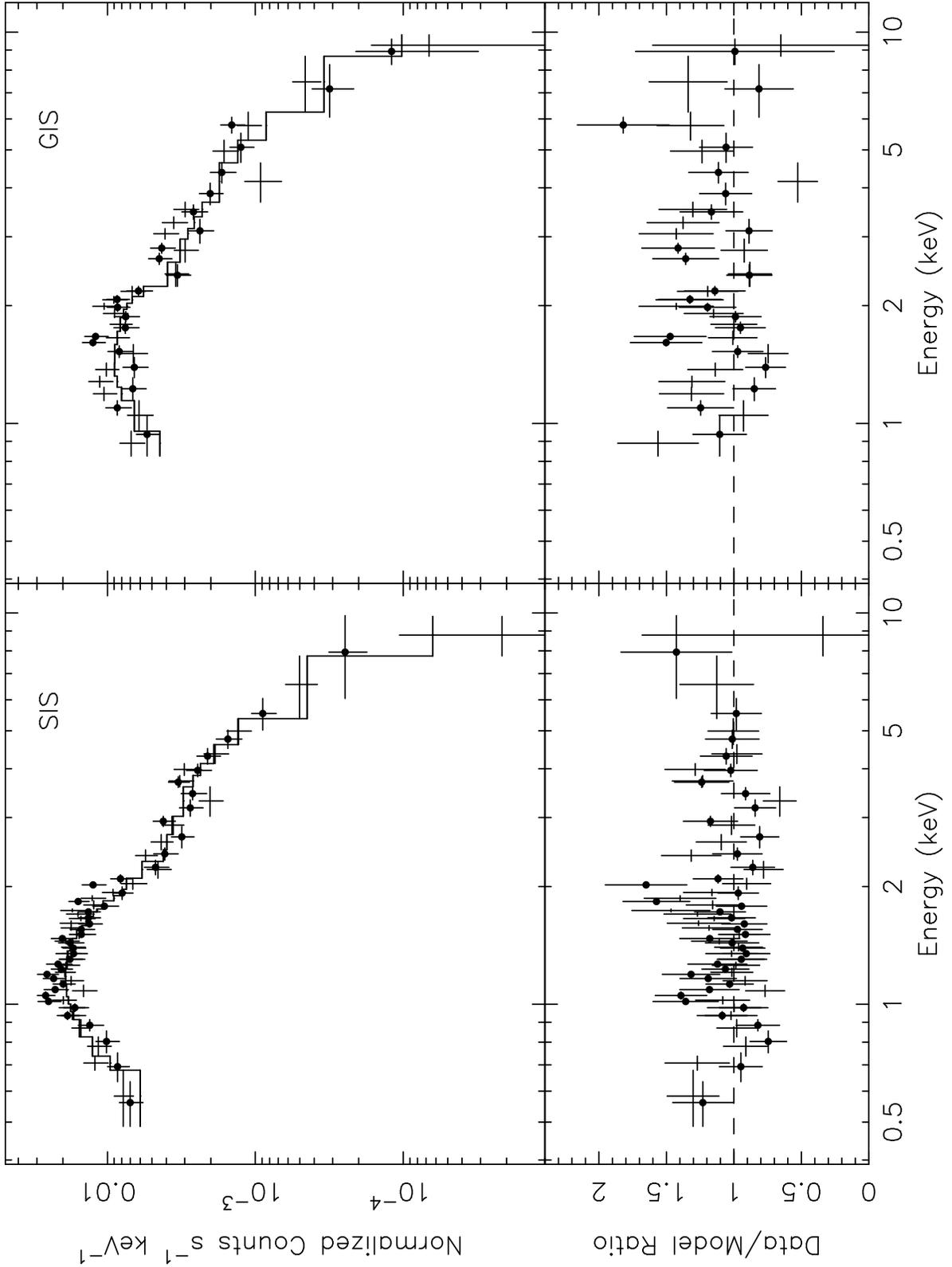}
\caption{{\small{The combined {\it ASCA} SIS spectrum ({\it left}) and the GIS 
spectrum ({\it right}).  The top panels shows the normalized counts of the 
two observations, where data from the first (1997 December 17 UT) and second 
(1997 December 23 UT) observation are shown as unmarked and marked points, 
respectively.  The histograms denote the best-fitting absorbed power-law 
model.  The ratio of the data and the model is illustrated in the bottom 
panels.  No significant residuals are seen.}}}
\end{figure}

\begin{deluxetable}{lllcccc}
\tablecolumns{7}
\tablewidth{6.9in}
\small
\def\h{\hskip -3 mm}
\def\ha{\hskip -2 mm}
\def\hb{\hskip +5 mm}
\tablenum{1}
\tablecaption{SPECTRAL FITS FOR NGC 4639}
\tablehead{
\colhead{\h Model} &
\colhead{$N_{\rm H}$} &
\colhead{$\Gamma$} &
\colhead{\h $F$(0.5--2 keV)} &
\colhead{\ha $F$(2--10 keV)} &
\colhead{\h $f_{\rm thermal}$} &
\colhead{$\chi^2_{\nu}$/dof} \nl
\colhead{\h } &
\colhead{(10$^{20}$ cm$^{-2}$)} &
\colhead{} &
\colhead{\h (\flux)} &
\colhead{\ha (\flux)} &
\colhead{\h (\%)} &
\colhead{}
}
\startdata
{\h \underbar {\it ASCA}          } &                 &                   &             &              &              &            \nl
\h Power Law           \dotfill     & 7.3 (2.2--12.9) & 1.68 (1.56--1.80) &\h 4.7\e{-13}&\ha 1.1\e{-12}&\h \nodata    &\h 0.98/491\nl
\h Power Law + Thermal \dotfill     & 6.1 ($<$18.4)   & 1.67 (1.52--1.84) &\h 4.7\e{-13}&\ha 1.1\e{-12}&\h $<$25      &\h 0.98/489\nl
\h                                  &                 &                   &              &               &            &            \nl
\h {\underbar {\it ASCA}} {\underbar and} {\underbar {\it ROSAT}} &&&&&& \nl
\h Power Law           \dotfill     & 2.0 (1.2--3.3)  & 1.58 (1.51--1.66) &\h 2.5\e{-13}&\ha 1.1\e{-12}&\h \nodata    &\h 0.99/504\nl
\h Power Law + Thermal \dotfill     & 1.9 ($<$17.9)   & 1.58 (1.48--1.68) &\h 2.5\e{-13}&\ha 1.1\e{-12}&\h $<$20      &\h 0.99/502\nl
\enddata
\tablecomments{The errors quoted are at the 90\% level for two interesting parameters.}
\end{deluxetable}

\begin{deluxetable}{llc}
\tablecolumns{3}
\tablewidth{4.0in}
\def\h{\hskip -3 mm}
\def\ha{\hskip -2 mm}
\def\hb{\hskip +5 mm}
\tablenum{2}
\tablecaption{X-RAY FLUX HISTORY OF NGC 4639}
\tablehead{
\colhead{Detector \ \ \ \ \ \ \ \ \ \ \ \ \ \ \ \ } &
\colhead{Dates} &
\colhead{$F$(0.5--2 keV)} \nl
\colhead{} &
\colhead{} &
\colhead{(\flux)}
}
\startdata
{\it Einstein} IPC \dotfill & 1980 Jun. 27 & 4.2\e{-13} \nl
{\it ROSAT} HRI    \dotfill & 1992 Jun. 27 & 8.0\e{-13} \nl
{\it ROSAT} HRI    \dotfill & 1992 Dec. 12 & 4.0\e{-13} \nl
{\it ROSAT} PSPC   \dotfill & 1993 Jun. 30 & 2.5\e{-13} \nl
{\it ASCA}         \dotfill & 1997 Dec. 17 & 4.6\e{-13} \nl
{\it ASCA}         \dotfill & 1997 Dec. 23 & 5.0\e{-13} \nl
\enddata
\end{deluxetable}

Previous {\it ASCA} observations have shown that low-luminosity AGNs typically 
emit a soft, thermal component in their X-ray spectra (e.g., Serlemitsos 
\etal 1996; Ptak \etal 1999), and it would be of interest to see if NGC 4639 
follows this pattern.  We attempted to fit the spectra with a model consisting 
of a thermal, Raymond-Smith (1977) plasma plus a power-law component.  Since 
the statistics are limited in this case,  we fixed the temperature and the 
abundance of the plasma to the mean values found in Ptak et al. (1999), namely 
$kT$ = 0.7 keV and $A$ = 0.043 $A_{\odot}$.  This fit yielded an upper limit 
to the thermal flux of $4.4 \times 10^{-13}$ \flux, or $\sim$25\% of the total 
flux in the 0.5--2.0 keV band.

There is no evidence for significant Fe~K emission at 6.4 or 6.7 keV, the 
expected line energies for neutral and He-like iron, although the statistics 
are limited.  Assuming the feature to be Gaussian in shape, we find upper 
limits to the equivalent width of $\sim$800 eV for a narrow line ($\sigma = 
10$ eV) and $\sim$1000 eV for a broad line ($\sigma = 100$ eV).

In order to obtain a more sensitive measure of the absorbing column, we 
attempted to fit the {\it ASCA} data simultaneously with the {\it ROSAT} 
Position-Sensitive Proportional Counter (PSPC) spectrum taken by Koratkar 
\etal (1995).  The PSPC spectrum was extracted using a circular aperture of
diameter 1\farcm25.  The combined fit yields a photon index of $\Gamma$ 
= 1.58 (1.51--1.66), similar to that obtained from the {\it ASCA} data 
alone, but the 0.5--2.0 keV flux is lower (by $\sim$50\%), as is the column 
density, $N_{\rm H} = 2.0 (1.2-3.3) \times 10^{20}$ cm$^{-2}$, although the 
latter is statistically consistent with the {\it ASCA} fit.  A two-component 
(thermal plasma plus power law) fit once again limits the possible 
contribution from  a thermal component to \lax 20\%, and the parameters of the 
power-law component remain essentially unchanged.  Table 1 summarizes the 
results of the spectral fits.  Note that the slope we derive for the 
X-ray continuum is significantly flatter than the value of $\Gamma$ = 
2.27$\pm$0.40 determined by Koratkar \etal (1995) from fitting the PSPC data
alone.  In our joint fit we also do not find spectral residuals 
at 1 keV, as suggested by Koratkar et al.  These disagreements are not 
surprising given the limited PSPC bandpass and the low signal-to-noise ratio 
of the PSPC data.  Several authors have found discrepancies between {\it ASCA} 
and {\it ROSAT} PSPC results in the bandpass where they overlap (0.5--2.0 keV; 
see Iwasawa, Fabian, \& Nandra 1999 and references therein), which suggests 
that there exists a calibration problem, most likely attributed to the PSPC, 
on the order of 20\%--40\%.  Moreover, as discussed further below (\S\ 3.3), 
the nucleus of NGC 4639 clearly varies in the soft X-ray band.   The results 
of our simultaneous fit should, therefore, be treated with some caution, 
although including the PSPC data has a very small impact on the observed hard, 
power-law component.

\subsection{Variability}

\begin{figure}
\plotone{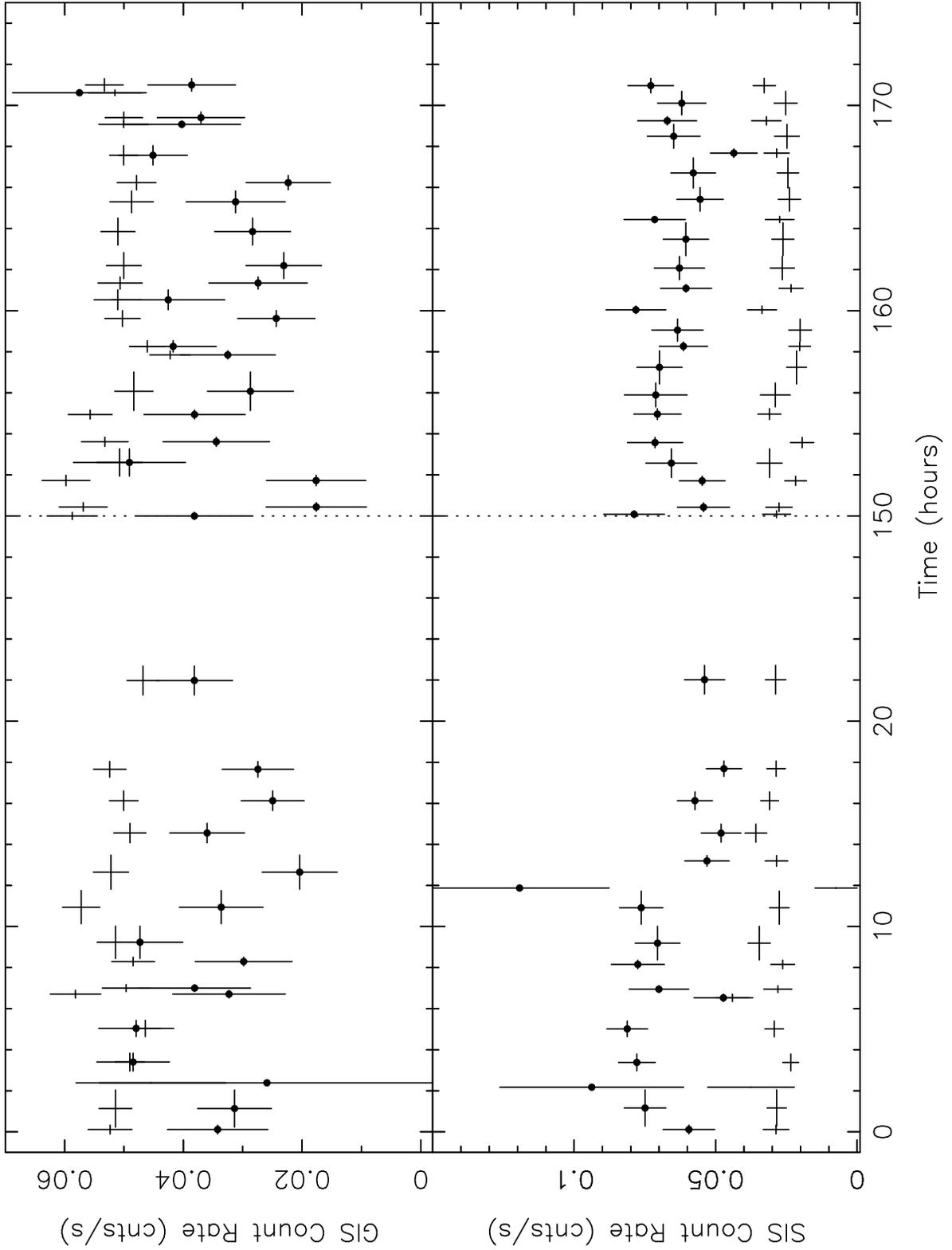}
\caption{GIS ({\it top}) and SIS ({\it bottom}) light curves of NGC 4639.  We 
have displayed the observations from both sessions on the same time axis, 
separated by the vertical dotted line.  The data are plotted as marked points, 
and the background level is shown as unmarked points, binned to a bin size of 
5760~s.}
\end{figure}

\begin{figure}
\plotone{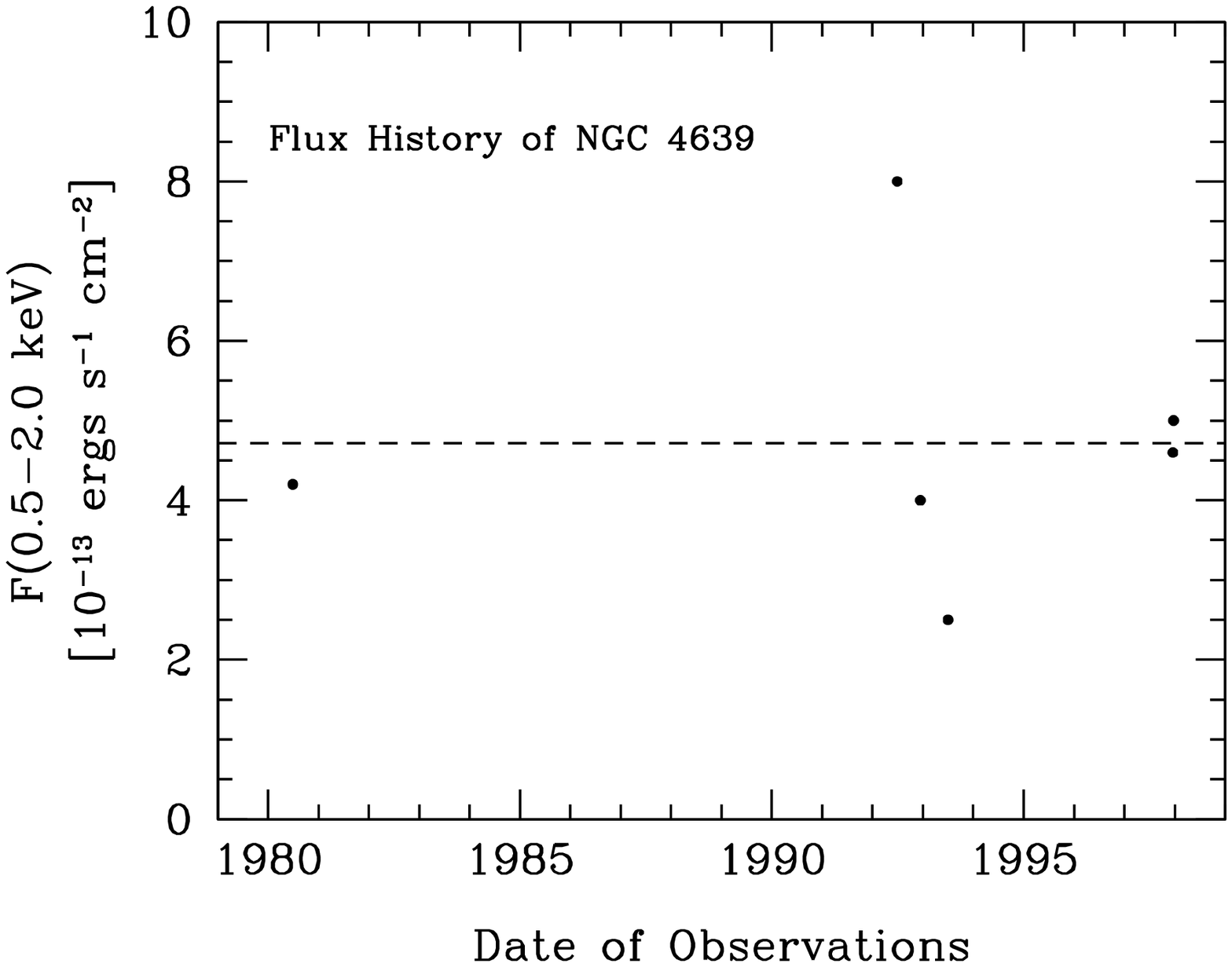}
\caption{The long-term X-ray light curve of NGC 4639 over the 0.5--2.0 keV
band.  The horizontal dashed line denotes the average flux of all the
measurements, which are listed in Table 2.}
\end{figure}

We searched for short-term flux variability within each of the two separate 
{\it ASCA} observations.  The top panel of Figure 3 shows the light curves 
obtained from combining the GIS2 and GIS3 data using a bin size of 5760~s, the 
length of one satellite orbit, and a similar plot for the SIS0 and SIS1 data 
is given in the bottom panel of the figure. To access the level of variability 
we performed a $\chi^2$ test to evaluate the null hypothesis that the flux is 
constant.  In the December 17 observation, variations are apparent at the 
level of $\Delta I/I\,\approx$ 50\% for the SIS data; a constant model fit 
produces $\chi^2_{\nu}$ = 2.8 for 15 degrees of freedom, and therefore can be 
rejected at a significance level of 99.9\%.  The background count rate 
stayed constant.  On the other hand, the SIS data for the second observation, 
taken six days later, showed no signs of variability.  A constant model 
fit yields $\chi^2_{\nu}$ = 0.8 for 21 degrees of freedom.   Marginal 
evidence of mild variability, significant at level of $\sim$95\%, may be 
present in both GIS observations.

We searched for variability between the two observations and found none.
To assess the nature of possible X-ray variability on longer timescales, we 
have retrieved archival data taken with the {\it Einstein} IPC in 1980, 
as well as data acquired on three epochs between 1992--1993 with the {\it 
ROSAT} HRI and PSPC.  For each of the {\it Einstein} and {\it ROSAT} 
instruments, we simulated the flux expected in the 0.5--2.0 keV band from the 
observed count rates and an assumed power-law model with $\Gamma$ = 1.58 and 
$N_{\rm H}$ = 2.0\e{20} cm$^{-2}$, the best-fit parameters obtained from the 
joint {\it ASCA} plus {\it ROSAT} PSPC fit (Table 1).  A summary of the flux 
history is presented in Table 2 and in Figure 4.  The source clearly undergoes 
large-amplitude flux changes over timescales of several months to a year.  In 
an interval of six months the two HRI pointings showed a factor of 2 
variability, and the largest peak-to-peak flux change is a factor of 3.2 in 
one year.  This type of long-term X-ray variability is commonly seen in other 
low-luminosity AGNs (Serlemitsos \etal 1996, and references therein).  
The variability could be intrinsic to the source, or it could be caused by 
variable absorption.

\section{Discussion}

\subsection{The X-ray Characteristics}

We have established that most of the X-ray emission in NGC 4639, both at hard
and soft X-ray energies, is nuclear.  Although the rather coarse angular
resolution of {\it ASCA} and {\it ROSAT} does not provide a stringent limit on
the physical source size, the morphology does suggest that most of the
emission originates from a compact region spatially coincident with the
optical center of the galaxy.  Flux variability, evident in the soft and
hard bands, further supports this conclusion.   Most of the X-ray emission
from NGC 4639, therefore, is likely to be directly associated with its weak
AGN observed at optical wavelengths.

The broad resemblance of NGC 4639 to luminous Seyfert 1 nuclei extends to the
spectral domain, despite the fact that its 2--10 keV luminosity is only
$\sim$8\e{40} \lum, at least a factor of 10--10$^3$ smaller than in most
Seyfert galaxies conventionally studied.  The spectrum of NGC 4639 from 0.5 to
10 keV, a moderately absorbed power law with a photon index of $\Gamma\,
\approx$ 1.6--1.7, is quite similar to the 2--10 keV spectra of more powerful
AGNs (Mushotzky 1982; Turner \& Pounds 1989; Nandra \etal 1997b).   Like other
broad-lined Seyfert nuclei the amount of obscuration seen toward the nucleus 
of NGC 4639 is small.  The absorbing column derived from the best-fitting 
model is consistent with pure foreground absorption from the Galaxy and at 
most requires a slight excess column from the host galaxy itself.  The most 
salient feature of the X-ray spectrum of NGC 4639 --- a ``canonical'' 
continuum slope --- is generally seen in {\it ASCA} observations of other
low-luminosity AGNs (see references cited in \S\ 1).  The limited photon
statistics of our data, unfortunately, prevent us from making a meaningful
assessment of spectral features that might be present.  We cannot address, for
instance, whether signatures of X-ray reprocessing by cold material, such as
the fluorescent Fe~K line or the Compton reflection bump, are present in the
spectrum.   The upper limit for the Fe~K line that we derive is large enough
to accommodate the range of observed line strengths.

The long-term X-ray flux history of NGC 4639 exhibits pronounced variability
on timescales of months to years.  With less confidence, we have argued
that it also plausibly experiences mild flux changes on intervals as short as
hours in the {\it ASCA} band.  The short-term lightcurves of low-luminosity
AGNs so far studied with {\it ASCA} are noteworthy for their unremarkableness
(Ptak \etal 1998).  Among more luminous AGNs, the amplitude of variability in
the hard X-ray band typically increases with decreasing X-ray luminosity (e.g.,
Barr \& Mushotzky 1986; Lawrence \& Papadakis 1993; Nandra \etal 1997a).  The
analysis of Nandra \etal (1997a) shows, for example, that a source such as
NGC 4051, with $L_{\rm X}$(2--10 keV) $\approx\,7 \times 10^{41}$ \lum\
(Mihara \etal 1994, assuming $d$ = 17 Mpc), has a variability amplitude
(``variance'') $\sim$25 times larger than, say, NGC 5548, whose
$L_{\rm X}$(2--10 keV) $\approx\,2 \times 10^{43}$ \lum\ (Mushotzky \etal
1995, assuming $d$ = 67 Mpc).  The objects in the sample of Ptak \etal (1998)
are at least an order of magnitude less luminous than NGC 4051, and yet they
vary much less than NGC 4051 does.  Their variability behavior clearly
deviates from the trend established by more luminous sources.  In the case of
NGC 4639, however, we have verified through simulations that variability at
the level of NGC 4051 cannot be excluded by the present data.

\subsection{The Nature of Central Source}

What is the production mechanism of the X-ray emission in NGC 4639?  Does it 
arise predominantly from processes similar to those in more luminous AGNs, or 
does it instead come from stellar sources such as X-ray binaries or 
starburst activity?  The strength of the broad component of the hydrogen 
Balmer lines in luminous Seyfert 1 nuclei scales with the observed X-ray 
luminosity in the 2--10 keV band (Elvis \etal 1978; Ward \etal 1988), a trend 
which can be interpreted as evidence that the optical emission lines are 
produced primarily through photoionization by the central continuum source.  
A similar correlation is seen in the soft X-ray band (Kriss, Canizares, \& 
Ricker 1980), and Koratkar \etal (1995) have shown that low-luminosity sources, 
including NGC 4639, also obey the correlation.  Terashima (1999) has recently 
found that broad-lined low-luminosity Seyferts and LINERs extend the 
$L_{{\rm H}\alpha}$--$L_{\rm X}$(2--10 keV) relation of Ward \etal (1988) 
toward the faint end.  This result suggests the hard X-ray emission in both 
luminosity classes share a common origin.  We find that NGC 4639, too, falls 
on the correlation.  The value of $L_{\rm X}$(2--10 keV)/$L_{{\rm H}\alpha}$ 
ranges between 2.7 to 6.1 depending on the assumed extinction affecting the 
optical emission (see below), well within the scatter of the correlation 
reported by Terashima (1999).

Perhaps a more stringent test can be made by examining the shape of the
ionizing continuum.  Although we do not have spectroscopic measurements in 
the ultraviolet (UV), nor do we have any constraints on the strength of the 
continuum in the extreme-UV band, we can calculate a rudimentary two-point 
spectral index between 2500 \AA\ and 2 keV, $\alpha_{\rm ox}$.  This quantity 
has an average value of 1.4 in quasars and 1.2 in luminous Seyfert 1s 
(Mushotzky \& Wandel 1989).   The central region of NGC 4639 has been observed 
with the Wide-Field Planetary Camera 2 (WFPC2) on {\it HST} through the 
F218W filter ($\lambda_c\,\approx$ 2150 \AA, FWHM $\approx$ 356 \AA); these 
observations will be reported by Calvani et al. (in preparation).  The nucleus
was clearly detected as a point source with a flux density at 2500 \AA\ 
of $f_{\nu}$ = 2.0\e{-28} \flux\ Hz$^{-1}$.  The true strength of the UV 
emission, however, is quite uncertain because of the possible effects of dust 
extinction.  As discussed in \S 3.2, the Galaxy contributes $E(B-V)$ = 0.04 mag
to the reddening, and an additional 0.03 mag, as determined from the decrement
of the narrow Balmer lines, may be internal to NGC 4639.   For the Galactic 
extinction law of Cardelli, Clayton, \& Mathis (1989), dereddening by $E(B-V)$ 
= 0.07 mag increases the 2500 \AA\ flux by a factor of $\sim$1.6, and we 
find $\alpha_{\rm ox}$ = 0.95$\pm$0.01, where the error bars reflect the 
range allowed for the X-ray slope.   If, on the other hand, we suppose that 
the value of $N_{\rm H}$ [$7.3 (2.2-12.9)\times10^{20}$ cm$^{-2}$] obtained 
from the {\it ASCA} data provides a better indicator of the absorbing column 
affecting the UV emission, then $E(B-V)$ = 0.13 (0.04--0.22) mag, and 
$\alpha_{\rm ox}$ = 1.01 (0.92--1.12).  Finally, we note that it is possible 
that the UV continuum experiences even more severe extinction than is 
indicated either by the narrow-line region Balmer decrement or by the 
X-ray absorbing column.  The relative intensities of the broad Balmer lines, 
measured  from the spectra published in Ho \etal (1995), suggest that the 
broad-line region is significantly reddened.  The observed ratios of 
H\al:H\bet:H$\gamma$ = 4.7:1.0:0.38.  Whereas radiative transfer effects and 
collisional excitation in the broad-line region render the H\al/H\bet\ ratio 
an unreliable reddening indicator (see, e.g., Netzer 1975; Kwan \& Krolik 
1981), the intrinsic H$\gamma$/H\bet\ ratio is less susceptible to variations 
in physical conditions.   For an intrinsic Case B$^{\prime}$ H$\gamma$/H\bet\ 
= 0.47, the observed value indicates $E(B-V)\,\approx$ 0.4 mag.  If this 
amount of reddening is appropriate for the continuum, and, furthermore, if 
the Galactic extinction curve is applicable, the UV emission would have to 
be increased by a factor of $\sim$14, and $\alpha_{\rm ox}\, \approx$ 1.3.  
Given the many uncertainties involved, it is obviously difficult to be 
definitive about the strength of the UV emission relative to the X-rays, 
but it appears to be roughly consistent with what is observed in 
more luminous Seyferts.  

As an additional consistency check, the observed luminosity of the Balmer 
lines can be compared with that predicted from recombination theory given the 
ionizing luminosity, which we can compute by interpolating the spectral 
energy distribution between the UV and X-ray bands.  For 
this comparison, we will use H\bet\ instead of H\al\ to avoid potential 
complications from collisional enhancement of H\al\ in the broad-line region.  
The total (narrow plus broad) luminosity of H\bet\, measured from the 
spectra in Ho \etal (1995), following the procedures outlined in Ho 
\etal (1997a, b), is 2.9\e{39} \lum, of which 94\% comes from the broad 
component.  Depending on the adopted reddening [$E(B-V)$ = 0.07--0.4 mag; see 
above], the corrected H\bet\ luminosity ranges from (0.4--1.1)\e{40} \lum.   
Case B recombination requires 8.5 Lyman continuum photons to generate an
H\bet\ photon (Osterbrock 1989), and so the corresponding ionizing photon rate, 
assuming a covering factor of unity, is $N_{\rm Lyc}$(H\bet) = 
(0.8--2.3)\e{52} \persec.  By contrast, the ionizing photon rate calculated by 
assuming an $f_{\nu}\,\propto\,\nu^{-\alpha}$ spectrum with $\alpha$ = 
$\alpha_{\rm ox}$ from the Lyman limit to 2 keV and $\alpha$ = 0.68 from 2 to 
10 keV, is $N_{\rm Lyc}$ = (1.5--7.4)\e{51} \persec, where the range once 
again reflects the range of probable extinction corrections.  Taken at 
face value, this simplistic calculation suggests that the observed high-energy
continuum falls short, by about a factor of 3 to 5, in supplying sufficient 
ionizing photons to sustain the observed line emission.  A possible solution
to this apparent discrepancy is to invoke additional extinction for the 
UV component.  Alternatively, the shape of the ionizing continuum through the 
unobserved extreme-UV region may be quite different from the simple power law 
that we assumed.  

\subsection{Comparison with NGC 5033}

It is instructive to compare the X-ray properties of NGC 4639 with those
of NGC 5033, another nearby, low-luminosity Seyfert 1 galaxy whose optical
spectrum is virtually indistinguishable from that of NGC 4639 (Ho \etal 1995,
1997b).  The {\it ASCA} spectrum of NGC 5033, recently analyzed by Terashima,
Kunieda, \& Misaki (1999), is well fitted with a lightly absorbed power law
with $\Gamma$ = 1.7, $N_{\rm H} = 9 \times 10^{20}$ cm$^{-2}$, and 
$L_{\rm X}$(2--10 keV) = 2.3\e{41} \lum\ (for an adopted distance of 18.7 
Mpc), to which is added a narrow Fe~K line with rest energy 6.4 keV.  With the 
exception of the iron line, which may have been detected in the spectrum of 
NGC 5033 but not in NGC 4639 because the former has significantly more counts, 
the X-ray spectral properties of the two objects are very similar.  As in 
NGC 4639, moderate variability on a timescale of $\sim$10$^4$~s was also found 
in NGC 5033.

Ho \etal (1997b) quote a broad H\al\ luminosity of 8.6\e{39} \lum.  After 
adjusting this to account for reddening, estimated to be 0.4 mag from 
the decrement of the narrow Balmer lines, we find 
$L_{\rm X}$(2--10 keV)/$L_{{\rm H}\alpha}$ = 8.8, a value once again 
in accord with the $L_{{\rm H}\alpha}$--$L_{\rm X}$(2--10 keV) relation 
of Terashima (1999).  We can perform the same exercise we did for NGC 4639 
concerning the UV emission, since NGC 5033 was also observed with 
{\it HST}/WFPC2, and it was detected as a compact UV source (Calvani et al., 
in preparation).  The observed flux density at 2500 \AA\ is $f_{\nu}$ = 
6.3\e{-27} \flux\ Hz$^{-1}$, which, when combined with the X-ray measurement, 
yields $\alpha_{\rm ox}$ = 1.2; including a reddening correction of 
0.4 mag increases $\alpha_{\rm ox}$ to 1.5.  In either case, the UV--X-ray 
slope lies securely within the range seen in luminous AGNs.   The 
uncorrected luminosity of the broad H\bet\ line, again determined from the 
data of Ho \etal (1995), is 1.8\e{39} \lum.  Even without extinction 
corrections, the Lyman continuum luminosity predicted from H\bet\ is only 
$\sim$10\%-40\% of the ionization budget estimated from interpolating the 
UV--X-ray continuum ($N_{\rm Lyc}$ = 1.0\e{52} and 4.1\e{52} \persec\ 
for $\alpha_{\rm ox}$ = 1.2 and 1.5, respectively), and including a reddening 
correction of 0.4 mag still brings the two estimates into comfortable 
agreement.  

\section{Summary}

Low-luminosity AGNs far outnumber ``classical'' Seyfert nuclei and quasars, 
but they remain a poorly understood class of objects.  Investigating sources 
such as NGC 4639, a relatively nearby low-luminosity Seyfert 1 galaxy, may 
provide valuable clues for understanding the AGN phenomenon in general and for 
exploring the connections between the ``active'' and ``normal'' galaxy 
populations.  The X-ray properties of NGC 4639 were previously discussed by 
Koratkar et al.  (1995) based on {\it ROSAT} observations.  The present study 
uses new {\it ASCA} data to extend the analysis into hard X-ray energies as 
well as UV measurements obtained with {\it HST}, and we reconsider the overall 
X-ray properties of this source in light of these new observations.

The nucleus of NGC 4639 was detected as a compact hard X-ray source.  The 
{\it ASCA} spectrum can be modeled as a slightly absorbed ($N_{\rm H} = 
7.3 \times 10^{20}$ cm$^{-2}$) power law with a photon index of 
$\Gamma\,\approx$ 1.7 and a luminosity of $L_{\rm X}$(2--10 keV) = 8.3\e{40} 
\lum.  Inspection of archival {\it Einstein} and {\it ROSAT} data reveals that 
the nucleus varies in the soft X-ray band on timescales of months to years, 
and there is marginal evidence of more rapid ($\sim$10$^4$~s) variability 
within the {\it ASCA} data sets.  If the variability is intrinsic to the 
source and not due to variable absorption, it would rule out the possibility 
that the power-law component arises from scattering, as was found for 
NGC 3147 (Ptak \etal 1996).  No significant contribution from a 
thermal plasma with a temperature of 0.7 keV, a trait common in 
other low-luminosity AGNs, was detected at soft X-ray energies.  

NGC 4639, despite being 1--3 orders of magnitude less luminous in the X-rays 
than most Seyfert galaxies previously studied, exhibits X-ray characteristics 
that are quite similar to those of more powerful sources.  Both the strength 
of its broad H\al\ emission line, measured from ground-based spectra, and 
its UV emission, measured from {\it HST} images, scale with the X-ray 
luminosity in roughly the same manner as do luminous AGNs.  There is a 
discrepancy of a factor of a few between the ionizing luminosity estimated 
from the UV--X-ray continuum and that inferred from the luminosity of the 
Balmer emission lines, but in view of the uncertainties inherent in the 
assumptions of our calculations, this disagreement is probably not serious.
Recent studies have noted key aspects in which low-luminosity AGNs 
differ from high-luminosity AGNs.  Ptak \etal (1998) find, for 
example, that the rapid variability behavior of low-luminosity objects
clearly departs from the trend set by luminous sources.  The variability 
amplitude of NGC 4639, on the hand, appears not to be anomalous.  It is 
also noteworthy that the UV--X-ray spectral energy distribution of NGC 4639, 
at least as crudely described by the $\alpha_{\rm ox}$ parameter, does not 
appear to be noticeably different from that of more luminous sources.  
This is in stark contrast to the set of low-luminosity objects studied 
by Ho (1999), most of which have X-ray luminosities comparable to that of 
NGC 4639, whose spectral energy distributions tend to be much more 
prominent in the X-rays compared to the UV band.  NGC 4639 --- along with 
NGC 5033, a source in many ways similar to NGC 4639 that we also briefly 
discuss --- serve as useful reminders that some low-luminosity AGNs appear
to be simple extensions of high-luminosity objects.

\acknowledgments
L.~C.~H. acknowledges partial support from NASA grants NAG 5-3556 and 
AR-07527.02-96A, the latter awarded by the Space Telescope Science Institute 
(operated by AURA, Inc., under NASA contract NAS5-26555).  Jane Turner, the 
referee, offered helpful suggestions for improving the paper.  We thank 
Humberto Calvani for supplying the {\it HST} UV flux measurements for NGC 4639 
and NGC 5033.


\clearpage

\vskip 0.5truein
\centerline{\bf{References}}
\medskip

\refindent 
Barr, P., \& Mushotzky, R.~F. 1986, \nat, 320, 421

\refindent
Bohlin, R.~C., Savage, B.~D., \& Drake, J.~K. 1978, \apj, 224, 132

\refindent
Cardelli, J.~A., Clayton, G.~C., \& Mathis, J.~S. 1989, \apj, 345, 245
 
\refindent 
de Vaucouleurs, G., de Vaucouleurs, A., Corwin, H.~G., Jr., Buta, R.~J.,
Paturel, G., \& Fouqu\'e, R. 1991, Third Reference Catalogue of Bright
Galaxies (New York: Springer)

\refindent 
Elvis, M., Maccacaro, T., Wilson, A.~S., Ward, M.~J., Penston, M.~V., Fosbury,
R.~A.~E., \& Perola, G.~C. 1978, \mnras, 183, 129

\refindent 
Filippenko, A.~V., \& Sargent, W.~L.~W. 1985, \apjs, 57, 503

\refindent 
Filippenko, A.~V., \& Sargent, W.~L.~W. 1986, in Structure and Evolution
of Active Galactic Nuclei, ed. G. Giuricin \etal (Dordrecht: Reidel), 21

\refindent
Gaskell, C.~M., \& Ferland, G.~J. 1984, \pasp, 96, 393

\refindent 
Ho, L.~C. 1999, \apj, in press (May 10 issue)

\refindent 
Ho, L.~C., Filippenko, A.~V., \& Sargent, W.~L.~W. 1995, \apjs, 98, 477

\refindent 
Ho, L.~C., Filippenko, A.~V., \& Sargent, W.~L.~W. 1997a, \apjs, 112, 315

\refindent 
Ho, L.~C., Filippenko, A.~V., Sargent, W.~L.~W., \& Peng, C.~Y. 1997b, \apjs,
112, 391

\refindent 
Ishisaki, Y., \etal 1996, PASJ, 48, 237

\refindent 
Iwasawa, K., Fabian, A.~C., \& Nandra, K. 1999, \mnras, in press 
(astro-ph/9904071)

\refindent 
Iyomoto, N., Makishima, K., Matsushita, K., Fukazawa, Y., Tashiro, M., \&
Ohashi, T. 1998, \apj, 503, 168

\refindent 
Koratkar, A.~P., Deustua, S., Heckman, T.~M., Filippenko, A.~V., Ho, L.~C., \&
Rao, M. 1995, \apj, 440, 132

\refindent 
Kriss, G.~A., Canizares, C.~R., \& Ricker, G.~R. 1980, \apj, 242, 492

\refindent 
Kwan, J., \& Krolik, J.~H. 1981, \apj, 250, 478

\refindent 
Lawrence, A., \& Papadakis, I. 1993, \apj, 414, L85


\refindent 
Mihara, T., Matsuoka, M., Mushotzkym R.~F., Kunieda, H., Otani, C., Miyamoto,
S., \& Yamauchi, M. 1994, \pasj, 46, L137

\refindent 
Murphy, E.~M., Lockman, F.~J., Laor, A., \& Elvis, M. 1996, \apjs, 105, 369

\refindent 
Mushotzky, R.~F. 1982, \apj, 256, 92

\refindent 
Mushotzky, R.~F., Fabian, A.~C., Iwasawa, K., Kunieda, H., Matsuoka, M.,
Nandra, K., \& Tanaka, Y. 1995, \mnras, 272, L9

\refindent 
Mushotzky, R.~F., \& Wandel, A. 1989, \apj, 339, 674

\refindent 
Nandra, K., George, I.~M., Mushotzky, R.~F., Turner, T.~J., \& Yaqoob, T.
1997a, \apj, 476, 70

\refindent 
Nandra, K., George, I.~M., Mushotzky, R.~F., Turner, T.~J., \& Yaqoob, T.
1997b, \apj, 477, 602

\refindent 
Netzer, H. 1975, \mnras, 171, 395

\refindent 
Nicholson, K.~L., Reichert, G.~A., Mason, K.~O., Puchnarewicz, E.~M.,
Ho, L.~C., Shields, J.~C., \& Filippenko, A.~V. 1998, \mnras, 300, 893

\refindent 
Osterbrock, D.~E. 1981, \apj, 249, 462

\refindent 
Osterbrock, D.~E. 1989, Astrophysics of Gaseous Nebulae and Active
Galactic Nuclei (Mill Valley: University Science Books)

\refindent 
Ptak, A., Serlemitsos, P.~J., Yaqoob, T., \& Mushotzky, R. 1999, \apjs, 120, 179

\refindent 
Ptak, A., Yaqoob, T., Mushotzky, R., Serlemitsos, P., \& Griffiths, R.
1998, \apj, 501, L37

\refindent 
Ptak, A., Yaqoob, T., Serlemitsos, P.~J., Kunieda, H., \& Terashima, Y. 1996,
\apj, 459, 542

\refindent 
Raymond, J.~C., \& Smith, B.~W. 1977, \apjs, 35, 419

\refindent 
Sandage, A., Saha, A., Tammann, G.~A., Labhardt, L., Panagia, N., \& Macchetto, 
F.~D. 1996, \apj, 460, L15

\refindent 
Sandage, A.~R., \& Tammann, G.~A. 1981, A Revised Shapley-Ames Catalog of
Bright Galaxies (Washington, DC: Carnegie Inst. of Washington)

\refindent 
Serlemitsos, P., Ptak, A., \& Yaqoob, T. 1996, in ASP Conf. Proc. 103,
The Physics of LINERs in View of Recent Observations, ed. M. Eracleous et al.
(San Francisco: ASP), 70

\refindent 
Tanaka, Y., Inoue, H., \& Holt, S.~S. 1994, \pasj, 46, L37

\refindent 
Terashima, Y. 1999, Adv. Space Res., in press 

\refindent 
Terashima, Y., Kunieda, H., \& Misaki, K. 1999, \pasj, in press (Vol. 52)

\refindent 
Turner, T.~J., \& Pounds, K.~A. 1989, \mnras, 240, 833

\refindent 
Ward, M.~J., Done, C., Fabian, A.~C., Tennant, A.~F., \& Shafer, R.~A.
1988, \apj, 324, 767

\refindent 
Weedman, D.~W. 1976, \apj, 208, 30

\end{document}